\documentclass{article}
\usepackage{spconf,graphicx}
\usepackage{enumitem}
\usepackage{float}
\usepackage{placeins}
\usepackage{amsmath}
\usepackage{amssymb}
\usepackage{bm} 
\usepackage{hyperref}
\usepackage{enumitem}
\usepackage{xcolor}

\title{SPC to 3D: Novel View Synthesis from Binary SPC via I2I translation}
\name{Sumit Sharma, Gopi Raju Matta, Kaushik Mitra}
\address{Indian Institute of Technology, Madras, India}
\begin{document}
%
\maketitle
\begin{abstract}
Single Photon Avalanche Diodes (SPADs) represent a cutting-edge imaging technology, capable of detecting individual photons with remarkable timing precision. Building on this sensitivity, Single Photon Cameras (SPCs) enable image capture at exceptionally high speeds under both low and high illumination. Enabling 3D reconstruction and radiance field recovery from such SPC data holds significant promise. However, the binary nature of SPC images leads to severe information loss, particularly in texture and color, making traditional 3D synthesis techniques ineffective. To address this challenge, we propose a modular two-stage framework that converts binary SPC images into high-quality colorized novel views. The first stage performs image-to-image (I2I) translation using generative models such as Pix2PixHD, converting binary SPC inputs into plausible RGB representations. The second stage employs 3D scene reconstruction techniques like Neural Radiance Fields (NeRF) or Gaussian Splatting (3DGS) to generate novel views. We validate our two-stage pipeline (Pix2PixHD + Nerf/3DGS) through extensive qualitative and quantitative experiments, demonstrating significant improvements in perceptual quality and geometric consistency over the alternative baseline.


\end{abstract}
\section{Introduction}
\label{sec:intro}
Single-photon camera pixel possesses remarkable sensitivity, capable of detecting individual photons and recording their arrival times with picosecond precision. Single Photon Cameras (SPCs), built on SPAD arrays, can operate in extremely low-light environments and at high temporal resolutions, making them highly attractive for scientific, industrial, and defense applications. However, these sensors typically output binary images due to photon detection thresholds, resulting in limited intensity and color information.

This binary representation poses a major obstacle for downstream vision tasks such as 3D reconstruction or novel view synthesis. Conventional approaches, such as Neural Radiance Fields (NeRF) or 3D Gaussian Splatting (GS), assume rich visual inputs with well-defined textures and lighting --- which SPC imagery fundamentally lacks.

To overcome this, we introduce a modular framework that first translates SPC binary inputs into RGB-like images using an image-to-image (I2I) model, and then reconstructs 3D scenes using NeRF or GS. This two-stage strategy allows independent optimization of the translation and reconstruction stages while enabling modular analysis of different model combinations. Our experiments reveal that this framework outperforms several existing pipelines on perceptual and geometric quality metrics. Moreover, we also present a physics-grounded SPC simulator based on Poisson statistics and dead-time modeling to facilitate reproducible training and evaluation.

Our contributions are threefold:
\begin{itemize}
\item We propose a novel application of I2I + NeRF/GS pipelines for processing binary SPC imagery.
\item We generated a training dataset using a statistically grounded SPC simulator that models photon detection based on physics principles.
\item We evaluate eight different I2I + 3D model combinations and highlight the effectiveness of Pix2PixHD + 3DGS/NeRF for high-quality 3D synthesis.
\end{itemize}

\section{Related Work}
\subsection{Image-to-Image translation}
Generative adversarial networks (GANs) \cite{goodfellow2020generative} are designed to model the distribution of natural images by generating samples that are indistinguishable from real-world images. They have enabled diverse applications, including image generation \cite{arjovsky2017wasserstein,radford2015unsupervised,zhao2016energy}. Generative image models are broadly categorized into parametric and non-parametric approaches. Non-parametric models synthesize images by matching patches from a database of existing images and have been widely used in texture synthesis \cite{efros1999texture}, super-resolution \cite{freeman2002example}, and image in-painting \cite{hays2007scene}. However, achieving photorealistic image generation remains challenging.
Adversarial learning has been extensively explored for image-to-image translation \cite{isola2017image}, where an input image from one domain is transformed into a corresponding output image in another domain using paired data. Unlike traditional L1 loss, which often leads to blurry outputs \cite{isola2017image}, adversarial loss \cite{goodfellow2020generative} has demonstrated superior effectiveness. Image-to-image translation problems are commonly formulated as per-pixel classification tasks \cite{long2015fully,xie2015holistically}, treating each output pixel as conditionally independent given the input image, thus considering the output space as "unstructured." In contrast, conditional GANs learn a structured loss \cite{isola2017image}, enabling more coherent and realistic synthesis.
Chen and Koltun \cite{chen2017photographic} proposed a method for generating photorealistic images conditioned on semantic layouts. Given a semantic label map, their approach synthesizes images that adhere to the input layout, functioning as a rendering engine that translates a two-dimensional semantic specification into a corresponding photographic representation. However, they highlighted the instability and optimization challenges of training conditional GANs for high-resolution image generation. While their method produced high-resolution results, the generated images often lacked fine details and realistic textures. To address these issues, Pix2PixHD \cite{wang2018high} introduced a novel objective function, multi-scale generators, and discriminators, significantly stabilizing training and yielding substantially improved high-resolution image synthesis.
\subsection{Radiance Fields}
Neural Radiance Fields (NeRF) \cite{mildenhall2020nerf} have revolutionized 3D vision with their ability to synthesize photo-realistic views using a neural implicit representation optimized via differentiable volume rendering. While subsequent works have enhanced NeRF's rendering quality and efficiency \cite{barron2021mip,muller2022instant}, 3D Gaussian Splatting (3DGS) \cite{kerbl20233d} has emerged as a faster alternative. By replacing NeRF’s ray marching with efficient rasterization, 3DGS achieves high-quality scene reconstruction with real-time rendering.
 
\section{Two Stage Approach for Novel View Synthesis}
\subsection{Single Photon Camera Simulator}
To train various Image-to-Image translation networks, we constructed a paired dataset comprising Single Photon Camera (SPC) images and their corresponding RGB color images. Despite continuous advancements in SPC technology and its functionalities, their availability in the commercial market remains limited. Therefore, we prepared a Single Photon Camera Simulator by leveraging insights from recent studies \cite{sharma2024transforming,goyal2021photon,liu2022single}. The simulator accepts color images as input flux, denoted as '$\phi$'. The SPC binary images are simulated on a channel-wise basis i.e., We simulated 1-bit per channel SPC images. The SPC simulator operates under the assumption that the SPC functions in passive mode. With each photon detection event, the Single-Photon Avalanche Diode (SPAD) undergoes a dead-time interval, during which it remains inactive and cannot detect additional photons. It is assumed that the arrival of photons per pixel follows Poisson statistics. Additionally, the number of detected photons '$N_{SPC}^{T}$' within a fixed exposure time '$T$' follows a renewal process, as detailed in Grimmett and Stirzaker’s work on probability \cite{grimmett2020probability}. Consequently, the mean and variance are calculated using the methodology outlined in Ingle and Proakis’s publication \cite{ingle2019high} as:

\begin{equation}\label{1}E[{N^{SPC}_{T}}] =  \frac{q_{SPAD}\phi T}{1 + q_{SPAD} \phi \tau_{d}}\end{equation}

\begin{equation}\label{2}Var[{N^{SPC}_{T}}] = \frac{q_{SPAD}\phi T}{(1 + q_{SPAD} \phi \tau_{d})^3},\end{equation}

where '$\phi$' represents the photon flux, i.e., the number of photons per second, and '$q_{SPAD}$' denotes the quantum efficiency of the SPC camera. Once the input image is loaded, it is processed by extracting the photon flux '$\Phi$', which represents the pixel intensity values. This photon flux is used to compute the mean photon count and variance photon count according to the equations outlined in \cite{liu2022single} i.e., equations \ref{1} and \ref{2}. During the simulation, parameters such as exposure time '$T$' and SPC pixel sensitivity '$q_{SPAD}$' are adjusted, while the dead-time '$\tau_{d}$' remains fixed. The objective is to ensure the visibility of all regions simulated by the SPC simulator. The parameter '$q_{SPAD}$' is set to 0.45, and the dead-time '$\tau_{d}$' is set to 150 ns during the simulation. After simulating the photon counts per pixel, the results are thresholded to create binary images: if the simulated photon count is greater than zero, the pixel is set to 255 (indicating photon detection); otherwise, it is set to 0 (indicating no photon detection). Finally, the binary images are generated per channel based on the estimated photon counts and combined to form the final colour binary image. 
\begin{figure*}[h!]
    \centering
    \vspace{-5mm}
    \includegraphics[width= 0.9\textwidth]{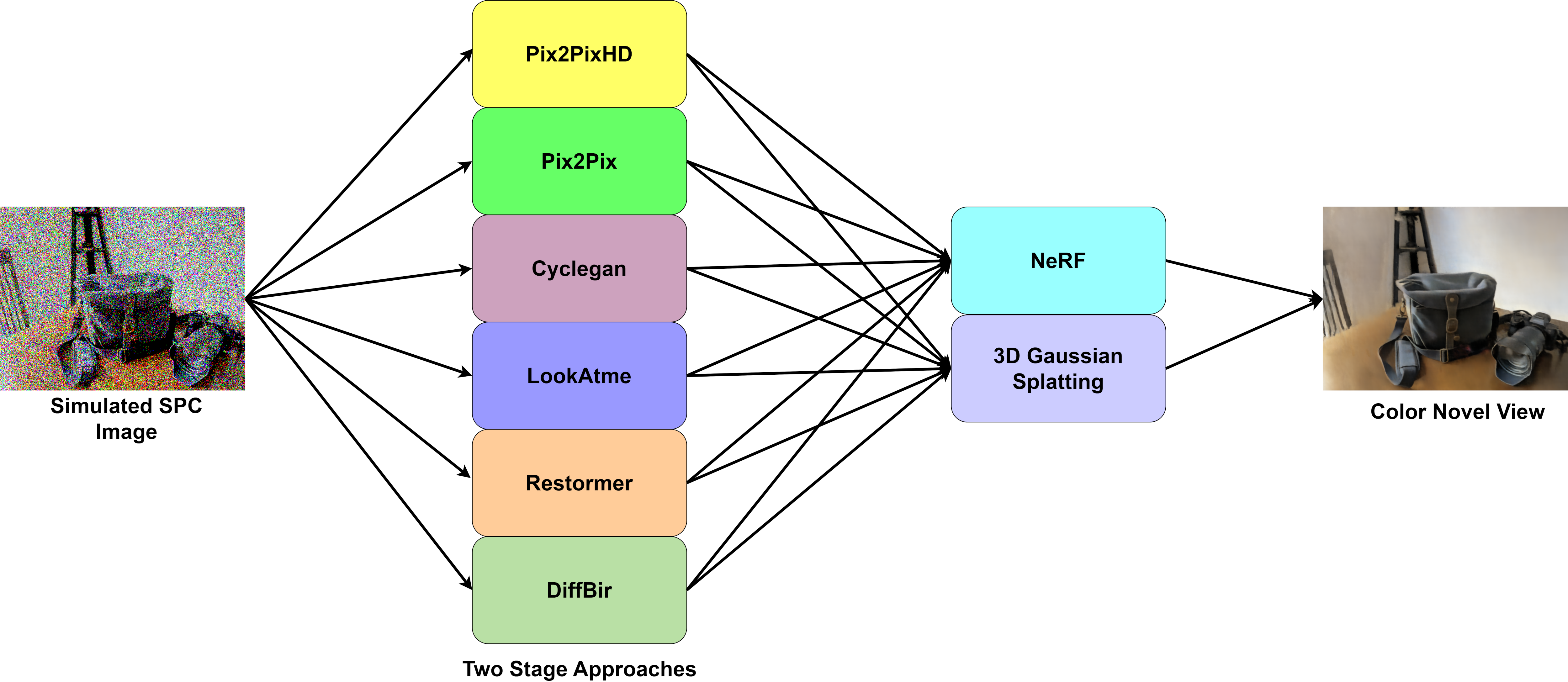}
    \caption{\textbf{Our proposed pipeline} illustrates possible two-stage approaches for transforming SPC binary images into color novel views. There are approximately eight possible combinations, among which Pix2PixHD + 3DGS/NeRF outperforms other state-of-the-art architectures.}
    \label{fig:Bin2nv}
\end{figure*}

\begin{figure*}[h!]
    \centering
    \vspace{-5mm} 
    \includegraphics[width=0.9\textwidth]{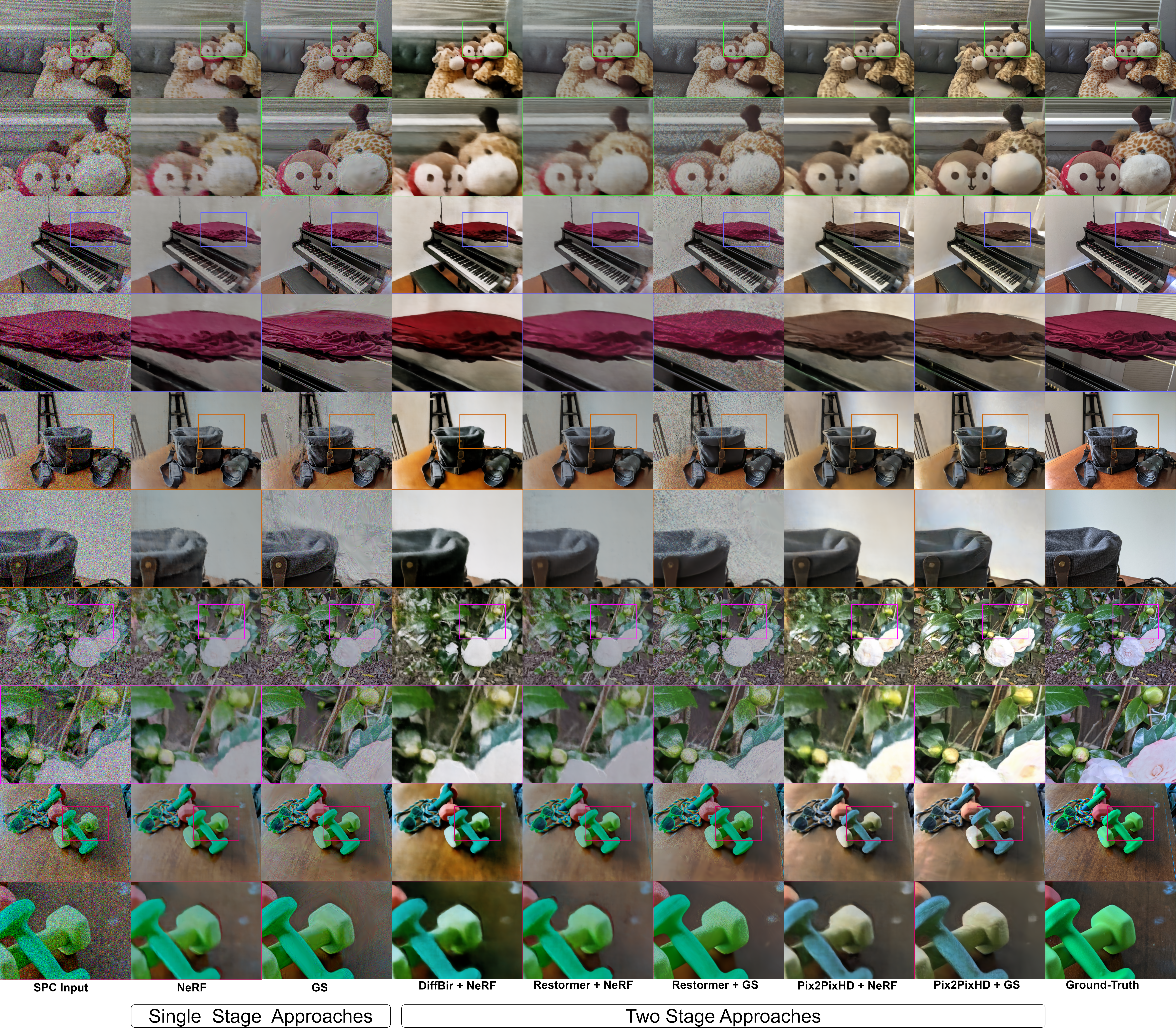}
    \caption{\textbf{Qualitative Results:} The first column represents the input simulated SPC images, while the last column depicts the ground truth images. The intermediate columns showcase the color novel views generated by various frameworks and their combinations. Upon examining the outputs, it can be observed that Pix2PixHD + 3DGS/NeRF produces the most accurate results in terms of texture detail and clarity. Please zoom in for a closer inspection.}
    \label{fig:Quality}
\end{figure*}

\subsection{Two-Stage Approach}
Since Single-Photon Camera (SPC) images are binary and lack visual appeal, generating color novel views directly from them is not feasible. To address this, an intermediate network is required to transform these unappealing binary images into visually meaningful color novel views.
To achieve this, we propose a two-stage approach:\newline
\textbf{Stage 1:} Transforming binary SPC images into color images using one of the following approaches:
\begin{enumerate}
    \item \textbf{Binary-to-Color Image Translation:} Directly using image-to-image translation networks to convert binary SPC images into color images.
    \item \textbf{Denoising Binary SPC Images:} First, applying a denoising step to enhance the quality of SPC images before translating them into color images.
\end{enumerate}
\textbf{Stage 2:} Transforming the generated colour images into colour novel views using NeRF and 3D Gaussian Splatting.\newline

In this study, we provide a comprehensive analysis of various two-stage approaches and identify the most effective ones. For Stage 1, which focuses on Binary-to-Color Image Translation, several image-to-image translation networks, including Pix2PixHD \cite{wang2018high}, Pix2Pix \cite{isola2017image}, CycleGAN \cite{zhu2017unpaired}, and LookatMe \cite{solano2023look}, were evaluated for performing the transformation between 3-channel simulated binary SPAD images and their corresponding color images. Additionally, in another task within Stage 1, aimed at Denoising Binary Images, we investigated the effectiveness of pretrained models from the Restormer \cite{zamir2022restormer} and DiffBir \cite{lin2025diffbir} architectures.

\section{Experiments}
To train the Image-to-Image translation networks, we curated a comprehensive training dataset leveraging the NeRF-LLFF dataset \cite{mildenhall2019local}. To enrich the diversity of the training data, we applied a series of image augmentations, including zooming, rotation, shearing, and flipping, which resulted in approximately 15,000 augmented images. Subsequently, SPC (Single-Photon Camera) images were simulated based on this augmented dataset to construct the final training set.For evaluation, we utilized the IBRNet-Collected-2 scenes \cite{zhou2021ibrnet}, where SPC images were similarly simulated. The trained models were then tested on this dataset to assess their performance.

In the Two-Stage approach, we translate simulated SPAD images into color images using various Image-to-Image translation networks, including \textbf{Pix2PixHD} \cite{wang2018high}, \textbf{Pix2Pix} \cite{isola2017image}, \textbf{CycleGAN} \cite{zhu2017unpaired}, and \textbf{LookatMe} \cite{solano2023look}. Among these, the \textbf{CycleGAN} and \textbf{LookatMe} architectures failed to successfully translate SPC binary images (3-channel). In contrast, the \textbf{Pix2PixHD} architecture was able to translate the images effectively. Additionally, we attempted to denoise the SPC binary images using a pretrained models of \textbf{Restormer} \cite{zamir2022restormer} architecture and also \textbf{DiffBir}\cite{lin2025diffbir} before transitioning to 3D rendering using NeRF and Gaussian splatting; however, these approaches proved inefficient. Therefore, within the Two-Stage framework, there are approximately \textbf{8} feasible working combinations possible.Please refer Fig \ref{fig:Bin2nv}.

\subsection{Implementation Details}
We trained various Image-to-Image translation networks using NVIDIA GeForce RTX 3090 GPUs. The NeRF model was trained for 50,000 iterations, while the Gaussian Splatting model was trained for 30,000 iterations. 
\subsection{Quantitative Results}
We conducted a quantitative analysis of both Single-Stage and Two-Stage approaches. The results are presented in Table \ref{tab:two_stage_results}. Notably, the combination of Pix2PixHD + NeRF/Gaussian-Splatting outperforms other state-of-the-art methods in terms of \textbf{PSNR}, \textbf{SSIM}, and \textbf{LPIPS}.

\begin{table}[H] 
\centering
\caption{PSNR, SSIM, and LPIPS Results for Two-Stage Approaches}
\label{tab:two_stage_results}
\begin{tabular}{|c|c|c|c|}
\hline
\textbf{Approach} & \textbf{PSNR $\uparrow$} & \textbf{SSIM $\uparrow$} & \textbf{LPIPS}$\downarrow$ \\ \hline
Pix2PixHD\cite{wang2018high}+NeRF\cite{mildenhall2020nerf} & \textbf{22.70} & \textbf{0.6843} & \textbf{0.4949} \\ \hline
Pix2PixHD\cite{wang2018high}+3DGS\cite{kerbl20233d} & \textit{22.21} & \textit{0.6772} & \textit{0.4251} \\ \hline
Pix2Pix\cite{isola2017image}+NeRF\cite{mildenhall2020nerf} & 18.35 & 0.5578 & 0.6473 \\ \hline
Cyclegan\cite{zhu2017unpaired}+NeRF\cite{mildenhall2020nerf} & 16.49 & 0.5377 & 0.6502 \\ \hline
Restormer\cite{zamir2022restormer} + NeRF\cite{mildenhall2020nerf} & 20.416 & 0.6562 & 0.4886 \\ \hline
Restormer\cite{zamir2022restormer} + 3DGS\cite{kerbl20233d} & 19.2733 & 0.4325 & 0.6063 \\ \hline
DiffBIR\cite{lin2025diffbir}+NeRF\cite{mildenhall2020nerf} & 19.353 & 0.6103 & 0.5009 \\ \hline
DiffBIR\cite{lin2025diffbir}+ 3DGS\cite{kerbl20233d} & 17.195 & 0.41825 & 0.47815 \\ \hline
\end{tabular}
\end{table}

\subsection{Qualitative Results}
We performed Qualitative analysis for both single stage and two-stage approaches. Please refer Fig \ref{fig:Quality}. Whereas in case of two-stage architecture, Pix2PixHD\cite{wang2018high}+NeRF\cite{mildenhall2020nerf} and Pix2PixHD\cite{wang2018high}+3DGS\cite{kerbl20233d} outperforms other state-of-the-art combinations.
 We observed that the improved adversarial loss proposed in Pix2PixHD \cite{wang2018high} plays a pivotal role in translating color information from SPC binary images to color novel views. This enhanced adversarial loss incorporates a feature-matching loss based on the discriminator, which stabilizes the training process by compelling the generator to produce natural image statistics at multiple scales. We also explored an approach where we first denoised the SPC images using state-of-the-art denoising models, such as Restormer \cite{zamir2022restormer} and DiffBIR \cite{lin2025diffbir}. However, this method resulted in rendered images that were lighter in color, lacked texture details, and exhibited increased noise. For visual reference, see Fig. \ref{fig:Quality}.

\subsection{Ablations with Single Stage Approaches}
In Single Stage approaches, we utilized two different frameworks to generate Color Novel Views from 3-channel Binary SPC images: the classic Neural Radiance Field (NeRF) \cite{mildenhall2020nerf} and 3D Gaussian Splatting \cite{kerbl20233d}. We trained both frameworks using 3-channel Binary SPC images and evaluated the models for novel view generation. While the classic NeRF framework exhibited slower training times and produced less impressive results in novel view generation, the faster Gaussian Splatting framework yielded better performance.
Both frameworks were directly applied to the Color Binary SPC simulated images. After training on these 3-channel Binary SPC images, we observed that the rendered images from NeRF were light in color, while the outputs from the 3D Gaussian Splatting framework displayed noticeable noise. Please refer Table \ref{tab:single_stage_results}, for quantitative analysis.From the quantitative analysis, we can infer that two-stage approaches yield better results compared to single-stage approaches, such as Pix2PixHD + GS/NeRF. Please refer Fig \ref{fig:Quality}.

\begin{table}[H] 
\centering
\caption{PSNR, SSIM, and LPIPS Results for Single-Stage Approaches}
\label{tab:single_stage_results}
\begin{tabular}{|c|c|c|c|}
\hline
\textbf{Approach} & \textbf{PSNR $\uparrow$} & \textbf{SSIM $\uparrow$} & \textbf{LPIPS}$\downarrow$ \\ \hline
NeRF\cite{mildenhall2020nerf} & 19.42 & 0.6252 & 0.50013 \\ \hline
3DGS\cite{kerbl20233d} & 18.643 & 0.5103 & 0.4855 \\ \hline
Pix2PixHD\cite{wang2018high}+NeRF\cite{mildenhall2020nerf} & \textbf{22.70} & \textbf{0.6843} & \textbf{0.4949} \\ \hline
Pix2PixHD\cite{wang2018high}+ 3DGS\cite{kerbl20233d} & \textit{22.21} & \textit{0.6772} & \textit{0.4251} \\ \hline
\end{tabular}
\end{table}
\section{Conclusion}
In this paper, we presented a modular two-stage framework that transforms binary SPC images into color novel views. By integrating Pix2PixHD for image translation with NeRF or Gaussian Splatting for 3D reconstruction, our method demonstrates notable improvements in texture detail and perceptual quality over other combinations. The use of a feature-aware adversarial loss, including feature-matching terms, contributes to enhanced synthesis quality. While the current architecture builds on existing components, its application to the challenging domain of binary SPC imagery highlights its practical relevance and effectiveness. Future extensions will explore unified, end-to-end models that jointly optimise both translation and reconstruction stages. In future work, we also plan to evaluate the pipeline on real SPC sensor data, which will further validate its robustness and demonstrate its practical viability in real-world imaging scenarios.

\bibliographystyle{IEEEbib}

\end{document}